\documentclass[aps,prx,twocolumn,showpacs,superscriptaddress]{revtex4-2}

\usepackage{graphicx}
\usepackage{dcolumn}
\usepackage{bm}
\usepackage{float}  
\usepackage{hyperref}

\newcounter{MD}
\newcommand*\MD%
{%
\ifnum\theMD>0%
{MD}\fi%
\ifnum\theMD=0%
{Molecular Dynamics (MD) \setcounter{MD}{1}}\fi%
}

\newcounter{SI}
\newcommand*\SI%
{%
\ifnum\theSI>0%
{SI}\fi%
\ifnum\theSI=0%
{Supplementary Information (SI) \setcounter{SI}{1}}\fi%
}

\begin{document}

\preprint{APS/123-QED}

\title{Patchy Particles Design: From Floppy Modes to Sloppy Dimensions}

\author{Gregory Snyder}
\author{Chrisy Xiyu Du}%
 \email{xiyudu@hawaii.edu}
\affiliation{%
 Mechanical Engineering, University of Hawai`i at Mānoa, Honolulu, HI, 96822
}%

\date{\today}

\begin{abstract}
Patchy particles have proven to be a prominent model for studying the self-assembly behavior of various systems, ranging from finite clusters to bulk crystal assemblies, and from synthetic colloidal particles to viruses. The patchy particle model is flexible, but it also comes with its own pitfalls---the potential design space is infinite. Many efforts have been put into building inverse-design frameworks that efficiently design patchy particles for targeted assembly behaviors. In contrast, little work has been done on investigating the interplay between different types of parameters that can be optimized for patchy particles, such as patch location, patch size, and patch binding energies. Here, by utilizing molecular dynamics with automatic differentiation, we elucidate the relationships between different potential optimization parameters and provide general guidelines on how to approach patchy particle design for various types of finite clusters. Specifically, we find that the design parameter landscape is highly dependent on the floppiness of the target structure, and we can identify stiff and sloppy parameters by computing the Hessians of all optimization parameters. 
\end{abstract}

\maketitle


\section{Introduction}
Patchy particles \cite{zhang2004self}, broadly defined as particles with discrete, attractive patches, are an excellent model system for studying the self-assembly behaviors of soft material systems, ranging from colloidal particles \cite{giacometti2010effects, sciortino2011reversible, rovigatti2018simulate, kalapurakal2023self} to proteins \cite{johnston2010modelling, klein2014studying, gnan2019patchy, li2025disorder, paine2025disassembly}. Known for their versatility, there are many realizations of patchy particles: triblock spheres that assemble into Hexagonal or Kagome lattice depending on the patch size \cite{chen2011directed, mao2013entropy}, patchy rhombuses and hexagons with programmable edges that assemble into both finite and bulk structures \cite{koehler2024particles, karner2025partially}, spheres with binding sites that assemble into Diamond structure \cite{zhang2005self, neophytou2021facile}, and triskelions that mimic the Clathrin protein assembly process \cite{den2010asymmetry}. Together with advances in theory and modeling of patchy particles, researchers have also made significant progress in patchy particle synthesis, such as functionalizing DNA patches on colloids \cite{feng2013dna} or vesicles \cite{liu2023organization}, colloidal fusion \cite{gong2017patchy}, and shape-shifting patchy particles \cite{zheng2017shape}. Other experimental materials platforms, such as magnetic handshake materials \cite{niu2019magnetic, liang2025magnetic}, DNA origami triangles \cite{hayakawa2022geometrically, hayakawa2024symmetry}, and even \emph{de novo} proteins \cite{li2023accurate, trubiano2024markov}, can be coarse-grained into patchy particle-like systems to understand their assembly behaviors. 

\begin{figure}[h!]
    \includegraphics[width = 0.9\linewidth]{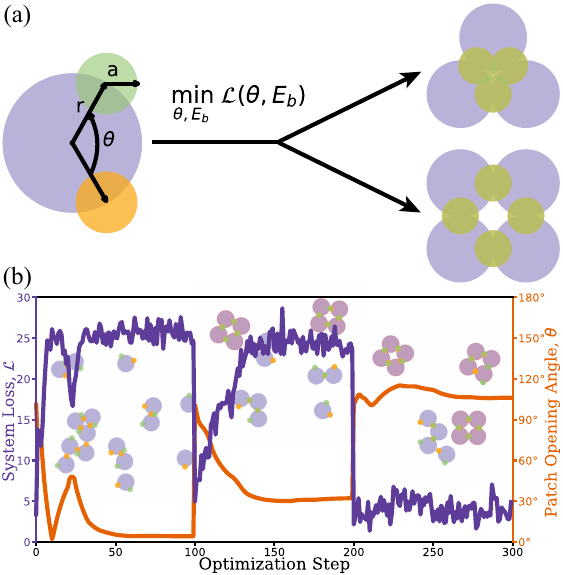}
    \caption{\textbf{System design} (a) We use a particular patchy particle model with one big sphere (radius $r$) as the body and two patches (radius $a$) A (green) and B (orange) as the attractive binding sites. The locations of the two patches are determined by the opening angle $\theta$, and the binding energies of the two patches are determined by the binding energy matrix $E_b$. We define a loss function $\mathcal{L}$ to find the best parameter combinations to assemble either a finite triangle or square robustly. (b) Parameter evolution for a typical self-assembly optimization with three different learning rates. At the beginning of every new learning rate, we select the best parameters from the previous learning rate as the starting point. The insert simulation snapshots show the final states of a sample batch. As the optimization converges, more squares start to form.}
    \label{fig:model}
\end{figure}

On the other hand, the versatility of the patchy particle model comes with a vast design space. There are many parameters that one could consider varying in a patchy particle system: the number of patches, the location of patches, the interaction between patches, and the shape of the particle. While the number of patches is discrete, all other parameters can be varied continuously, thus creating a high-dimensional design space. To effectively search through the giant design space for a patchy particle configuration with a targeted materials property in mind, various inverse-design strategies have been developed. Covariance matrix adaptation evolutionary strategy (CMA-ES) was employed to design the optimal patchy particle for finite cluster assembly \cite{long2018rational} and to optimize photonic band gaps for colloidal crystals \cite{ma2021inverse}. Neural Networks, combined with iterative learning, are used to design various self-assembly motifs in 2D patchy particle systems \cite{whitelam2021neuroevolutionary}. Boolean satisfiability problem (SAT) solvers were used to design patchy particles to better assemble the Diamond structure \cite{russo2022sat, beneduce2023engineering}, and Digital Alchemy with incorporated patchy interaction \cite{rivera2023inverse} was able to find designs for the Kagome lattice.

While these methods have been effective, one component that is hard to account for is the self-assembly kinetics. Many inverse-design methods are very good at finding a design that minimizes the free energy of the system, but do not guarantee a pathway for the system to successfully reach the targeted structure \cite{rivera2023inverse, wei2025designing}. To overcome this, inverse-design strategies that combine automatic differentiation and \MD\ can implicitly or explicitly account for self-assembly kinetics or pathways. Using JAX-MD \cite{schoenholz2019jax}, an end-to-end differentiable \MD\ engine, researchers were able to design colloidal cluster transition rates \cite{goodrich2021designing}, configurations for fast disassembly \cite{krueger2024tuning}, and pathways for work-minimization in non-equilibrium systems \cite{engel2023optimal}. With the recent addition of rotational degrees of freedom in JAX-MD \cite{king2024programming}, it can now perform inverse-design for patchy particles for any arbitrary parameter combination.

In this paper, we demonstrate how to use automatic differentiation to systematically investigate the parameter design space for patchy particles, rather than just finding the ``optimal" set of parameters. We employ a simple patchy particle model with two patches and optimize it for finite triangle and square assemblies. This approach enables us to uncover complex relationships among all potential design parameters, including patch size, patch opening angle, and patch binding energy. We provide detailed analysis to reveal the relationship between design parameters and physical properties (floppy vs. rigid) of the target structure. Lastly, we discuss a potential route to systematically examine the correlations between design parameters and identify stiff versus sloppy dimensions using Hessian information during the inverse design process.

\begin{figure}
    \centering
    \includegraphics[width = 1.0\linewidth]{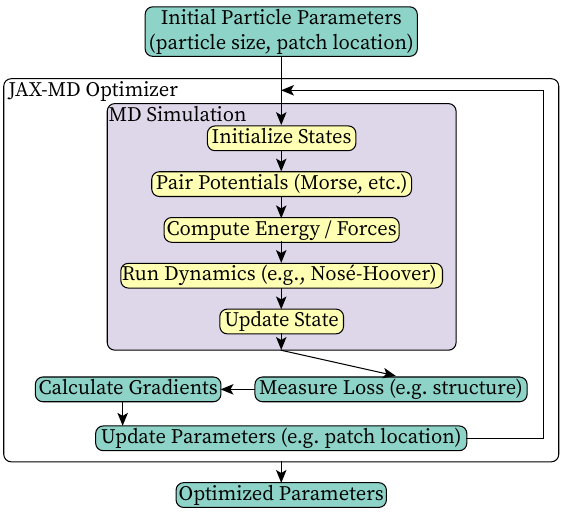}
    \caption{\textbf{JAX-MD optimization framework:} the flowchart shows how the parameters for optimization are threaded through the simulation, allowing us to retrieve the gradients after a complete forward \MD\ simulation.}
    \label{fig:protocol}
\end{figure}

\section{Method}

All the inverse-design and forward MD simulations in this paper are performed using JAX-MD \cite{schoenholz2019jax}. JAX-MD is an \MD\ engine that is end-to-end differentiable and GPU-accelerated. In particular, all our simulations utilize the \texttt{RigidBody} class developed in \cite{king2024programming}, allowing us not only to simulate anisotropic building blocks but also to directly take gradients on building block shapes. Fig.~\ref{fig:protocol} shows a flowchart of how gradients are being threaded through from the beginning of the \MD\ simulation to the end.

Our patchy particle model (see Fig.~\ref{fig:model}(a)) consists of three spheres of different species: a large center particle with two attractive patches (type A and B) along the surface. The locations of the two patches are determined by an opening angle $\theta$. The center particle determines the shape and size of the patchy particle, which is modeled using the soft sphere potential \cite{hansen_phase_1970}. We maintain a mass of 1.0, a radius of 1.0 ($\sigma=2.0$), $\epsilon_s = 1.0$ and $\alpha_s = 2.0$ of the center particle for all our simulations. The two attractive patches are massless, and the size of the patch is decided by the $\alpha$ parameter in the Morse potential \cite{morse_diatomic_1929}. Soft sphere potential and Morse potential are defined as follows:

Soft Sphere Potential:
\begin{equation}
 U_{s}(r) = \epsilon_s(\frac{\sigma}{r})^{\alpha_s}
 \label{eqn:softsphere}
\end{equation}

Morse Potential:
\begin{equation}
 U_{M}(r) = \epsilon_m(1-e^{-\alpha(r-r_0)})^2
 \label{eqn:morese}
\end{equation}

We explored ten different patch sizes initially for their assembly behaviors with $\alpha=[3, 4, 5, 6, 7, 8, 9, 10, 11, 12]$. These $\alpha$ values correspond to patch size ratios $a/r=[0.88, 0.66, 0.53, 0.44, 0.38, 0.33, 0.29, 0.26, 0.24, 0.22]$. Detailed conversions between $\alpha$ and $a/r$ can be found in the \SI. Parameter $\epsilon_m$ governs the binding strength between all the patches, and we will use $E_{AA}, E_{AB}$, and $E_{BB}$ to denote the interactions between patch types A and B.

In order to understand the relationship between patch size, opening angle, and binding energies for finite cluster assembly, we performed two different types of optimization runs: self-assembly and stability optimization. For self-assembly optimizations, we initialize the system with 15 or 18 particles (for triangles) or 16 particles (for squares). We initialize the system with all particles arranged randomly in a 2D box. For stability optimizations, we initialize the system with three particles (for triangles) or four particles (for squares). We deliberately arrange the particles with desired positions and randomized orientations. For both types of optimizations, we run the optimization for 300 steps using the \texttt{adam} optimizer \cite{adam} with three different learning rates: 0.5, 0.1, and 0.05, respectively, for 100 steps each. When we switch to the new learning rate, we find the parameters from the last 100 steps that yielded the most correct structures and use those values as the starting point for the new learning rate. For each optimization step, we ran forward simulations with a batch size of 16 for stability optimization and a batch size of 64 for self-assembly optimization, both with 40,000 or 60,000 timesteps and a time step of $dT = 1 \times 10^{-3}$ at a density of $\rho = 0.2$ and a temperature of $kT = 1.0$. Fig.~\ref{fig:model}(b) shows the parameter and loss value evolution for a typical optimization run. The number of batches was determined by performing benchmark simulations that took into consideration reducing gradient noise and computational time (see details in \SI).

\begin{figure}
    \centering
    \includegraphics[width = 1.0\linewidth]{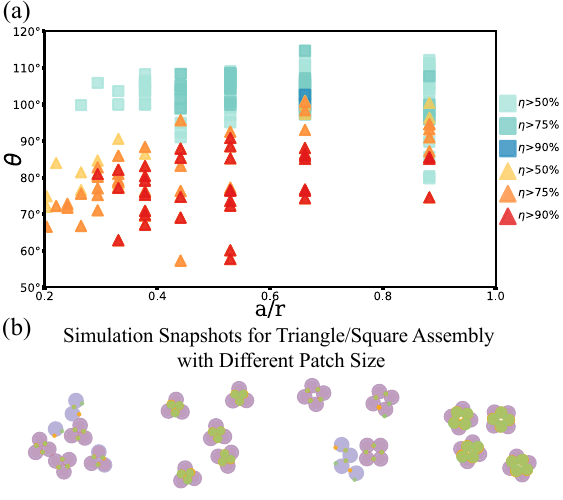}
    \caption{\textbf{Self-Assembly Optimization Results: }(a) Optimal patch opening angle with respect to different patch sizes for triangle and square assembly. The square markers indicate results for square assembly, while the triangle markers indicate results for triangle assembly. The color of the markers corresponds to the yield $\eta$ of a given optimization. We classify an optimization to be successful for $\eta>50\%$ and we plot three $\eta$ cutoffs: [50\%, 75\%, 90\%]. (b) Four sample snapshots of the forward simulation for the last step of the optimization. We note successful triangle and square formations for both small and large patch sizes.}
    \label{fig:selfassembly}
\end{figure}

Since the JAX-MD optimization process requires end-to-end differentiability, we cannot use a standard clustering algorithm as the loss function for the optimization. Instead, we match the local environment of each particle to a reference shape. Using square optimization as an example, we compute the three nearest neighbor distances of each particle in the final step of the simulation and calculate the relative distance difference compared to a perfect square. The final loss value is the sum of all the differences between all particles in the simulation. If the simulation formed all squares, the system loss should be minimized to 0. However, in most self-assembly simulations, not all particles can successfully form into a square, resulting in a loss that is significantly greater than zero.

Additionally, given the definition of the loss function, a final configuration of all wrong structures might result in a lower loss than a configuration of one square and 12 monomers. To ensure that the final optimized parameters will favor square formation, we apply a secondary measurement after each optimization step: using the clustering algorithm in freud \cite{freud2020} to compute the number of squares formed in each batch. We then compute the yield of the optimization as the number of batches with at least one square formed, divided by the total number of batches.

\begin{figure}[h!]
    \centering
    \includegraphics[width = 1.0\linewidth]{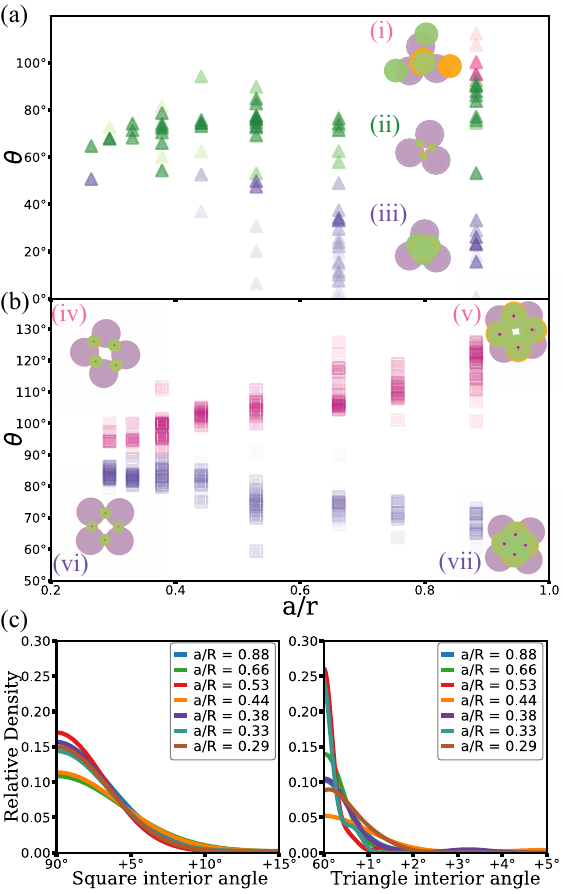}
    \caption{\textbf{Stabilization Optimization Results: }(a) Optimal patch opening angle with respect to different patch sizes for triangle stabilization. Each data point represents one successful stabilization optimization. The darker regions indicate multiple optimizations converged to the same patch opening angle. The data points are categorized into three colors: (i) pink, (ii) green, and (iii) purple, indicating three different stable triangle morphologies. (b) Optimal patch opening angle with respect to different patch sizes for square stabilization. Each data point represents a successful stabilization optimization. The darker regions indicate multiple optimizations converged to the same patch opening angle. The data points are categorized into two types: (iv-v) pink and (vi-vii) purple, indicating the two different square conformations. (c) Interior angle distribution for squares and triangles. While the interior angle distribution is agnostic of patch sizes, the angle distribution for squares is dramatically bigger than that of triangles.}
    \label{fig:stablization}
\end{figure}

\section{Result}

\textbf{Self-Assembly Optimization:} We first explored the effect of patch sizes on optimal patch opening angles for finite triangle and square self-assembly. In the previous study that introduced patchy particle optimization using JAX-MD \cite{king2024programming}, a single patch size was used during optimization. It was found that the particles preferred a patch opening angle bigger than the intuitive guess of $90^{\circ}$ for squares and $60^{\circ}$ for triangles. Here, we aim to investigate whether this result holds for other patch sizes or if there are other dependencies between the optimal patch size and opening angle.

Fig.~\ref{fig:selfassembly}(a) shows the relationship between patch opening angle and patch size ratios for both triangle and square assembly. The square markers show the optimization results for square assemblies of various patch sizes, while the triangle markers show the optimization results for triangle assemblies of various patch sizes. We note that there are different numbers of points for each patch size. The reason behind this is that we only plot the data that we label as successful here, and there are many optimization runs that either failed to converge or did not pass our success measurement bar. To measure success, we measure the yield ($\eta$) of triangles or squares during the optimization. $\eta$ is computed as the number of batches that had at least one square/triangle formed in the last step of optimization over the total number of batches. We consider an optimization successful when $\eta$ exceeds 50\%. In Fig.~\ref{fig:selfassembly}(a), we separate the data points by three different $\eta$ cut-offs: [50\%, 75\%, 90\%] and color them accordingly. We note that as the patch size decreases, fewer optimizations were successful, which is due to kinetic effects during the simulation. As the patch size decreases, the chance of two patches being near each other also decreases, leading to a potentially longer binding time scale. To reduce kinetic effects, we increased the forward simulation time from 40,000 steps to 60,000 steps, allowing us to obtain more data points. 

Looking at the relationship between patch opening angle and patch size, we notice two dramatically different trends for squares and triangles. As the patch size decreases, the optimal opening angle for triangles also decreases and approaches the intuitive guess of $60^\circ$. However, for square assemblies, we do not observe a similar trend. The optimal patch opening angles hover around $100^\circ$, and the only difference for different patch sizes is that the variance of optimal angles decreases as the patch size decreases. This result indicates that while optimizing for triangles and squares seems like the same type of inverse-design tasks, there is something fundamentally different between the two types of structures, causing different kinds of dependencies between the optimization parameters.

Lastly, we want to point out the ability of patch opening angles to select for different final assembled structures. Intuitively, as the patch size increases, the selective power of the patch opening angle will decrease, and no parameter combination can robustly assemble either squares or triangles. However, based on our results, we will lose the selective power only when the patch size is more than 60\% of the size of the body particle. This result informs us that as long as the patch opening angle is controlled, we can allow for a bigger patch to accelerate the assembly process. Additionally, the patch opening angle overlapping region also validates the robustness of our inverse-design method, as we can isolate the structure we are optimizing for, even though the patchy particle can form different polymorphs.

\textbf{Stabilization Optimization: }Since the correlation of patch opening angle and patch sizes differs significantly between triangle and square assemblies. We believe that the difference arises from the structural property differences between triangles and squares. Using our patchy particle model, the assembled triangle clusters are \emph{rigid}, while the assembled square clusters are \emph{floppy} \cite{mao2018maxwell}. To confirm our hypothesis, we performed stabilization optimizations for various patch sizes to see if the same trends hold compared to the self-assembly optimizations.

Fig.~\ref{fig:stablization}(a) shows the optimal opening angle for triangle stabilization of different patch sizes. From first glance, there is almost no trend between patch size and opening angle, apart from the increasing range of possible opening angles as patch sizes increase. We went through all the optimization simulations and were able to identify three different triangle morphologies (see Fig.~\ref{fig:stablization}(i)-(iii)) that are equivalent using our position-based loss function. Morphology (i) only occurs at the larger patch sizes, as it uses only one of the two patches as binding sites for the triangle cluster. Morphology (ii) is the desired configuration to have AB binding, while all six patches bind with each other in morphology (iii). 

Comparing the triangle stabilization results to the self-assembly ones, we notice that the trend of morphology (ii) generally agrees with the trend of triangle self-assembly. Morphology (i) is only a realistic configuration when no other particles exist in the system, as it is not self-limiting, and morphology (iii) is less kinetically favorable, as all patches need to lock into the same location. Combining the inverse-design results from both stabilization and self-assembly allows us to explore all the possible configurations and separate energy and kinetics contributions to the final optimized parameters.

Fig.~\ref{fig:stablization}(b) shows the optimal opening angle for square stabilization of different patch sizes. Unlike the triangle results, square stabilization results show a different trend. As patch sizes increase, a clear bifurcation appears for the optimal patch opening angle. If we extrapolate the trends of the patch opening angle, it will converge to $90^\circ$ as the patches reach a point particle. Fig.~\ref{fig:stablization}(iv)-(vii) show the configurations for square stabilization with high and low patch opening angles and small and big patch sizes. All four configurations fall under the same morphology with AB binding, the same as that of square self-assembly. We note that the stabilization results have a significant overlap with the self-assembly result for morphologies that appear in both sets of optimizations, and we provide a more detailed discussion in the \SI.

The stabilization optimizations allow us to explore all possible morphologies for triangle and square cluster assembly. While the two structures are relatively simple, we were able to identify seven different configurations that are hard to pinpoint with traditional forward \MD\ simulations. Comparing the morphologies between triangles and squares, we see that morphologies (i) and (iii) do not occur in square structures, as they require the structure to be space-filling. At the same time, we do not observe bifurcations in morphology (ii), as triangle structures are not floppy. 

To confirm that square structures are indeed floppy while triangle structures are not, we measure the interior angle fluctuation of square and triangle structures (see Fig.~\ref{fig:stablization}(c)) for different patch sizes in equilibrium \MD\ simulations. We see that \emph{floppiness} is an inherent structural property, thus independent of the building block properties (patch size and opening angles), and the distribution of square interior angles is much broader than that of triangles. 

\section{Discussion}

\begin{figure}[h!]
    \centering
    \includegraphics[width = 1.0\linewidth]{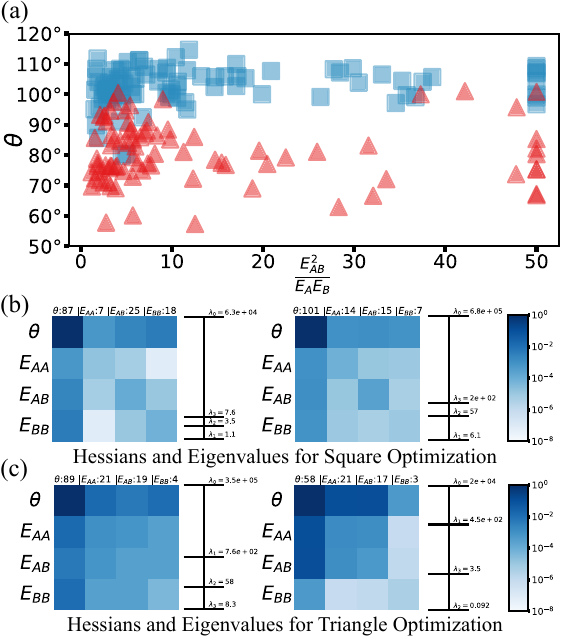}
    \caption{\textbf{Relationship Between Optimized Patchy Particle Parameters: }(a) Scatter plot of optimized patch opening angle as a function of on-target binding and off-target binding energy ratio ($E_{AB}^2/(E_{AA}E_{BB})$). Each data point represents one successful self-assembly optimization (square marker for square optimization and triangle marker for triangle optimization). The binding energy ratios are plotted with an upper limit of 50. (b-c) Hessians for the four optimized parameters (patch opening angle, $E_{AA}$, $E_{AB}$, and $E_{BB}$) and the corresponding eigenvalues for two different successful square and triangle optimizations. The optimized parameters are listed at the top of the Hessian matrices.}
    \label{fig:energy}
\end{figure}

By systematically investigating how the optimal patch opening angle depends on patch sizes for different self-assembled finite structures, we discovered that the correlations between potential design parameters can be drastically different and highly depend on the structural properties of the final assembled structure. Moreover, we demonstrate that gradient-based optimization tools such as JAX-MD not only can help us inverse-design ``optimal'' building block parameters for a given assembly task, but they can also serve as a tool to aid the understanding of parameter landscapes in a high-dimensional design space. 

In soft materials synthesis, oftentimes it is not possible to control all particle parameters precisely \cite{auer2001suppression, luo2019hierarchical, royall2024colloidal}. While many inverse-design methods provide a set of optimal parameters to achieve a specific material's assembly or behavior \cite{long2018rational, russo2022sat, rivera2023inverse, goodrich2021designing, king2024programming, wei2025designing}, it is also essential to discuss error tolerance. Our results not only discuss the possible single-parameter range that can lead to the same assembled structure, but also examine the coupled relationships between different parameters, such as patch size and patch opening angle. Our results can guide the synthesis of soft materials, whether it involves colloidal patchy particles \cite{gong2017patchy}, DNA-tethered particles \cite{feng2013dna}, or \emph{de novo} proteins \cite{li2023accurate}. For example, suppose one wishes to synthesize building blocks to self-assemble a rigid structure. In that case, it is crucial to have control over \emph{both the size and location} of the binding sites, as demonstrated in our triangle optimization results: the larger the patch size, the larger the opening angle. On the other hand, if one wishes to synthesize building blocks for floppy structure assembly, then the binding site \emph{location is the only important parameter}, and the size of the binding sites can be more forgiving. Having precise control over particles can be expensive, and our results can guide us on when low precision is good enough.

On the other hand, the parameters we chose to optimize are driven by the physical properties of soft material systems, and we do not know the coupling among these parameters \emph{a priori}. In previous work \cite{king2024programming}, it was demonstrated that the binding energies do not change significantly during optimization. In this work, our systematic investigations also result in the same conclusion (see details in \SI). However, after plotting optimized patch opening angle as a function of final binding energy ratio between on-target and off-target binding (see Fig.~\ref{fig:energy}(a)), we noticed that when the off-target binding energy ($E_{AA}$ and $E_{BB}$) was strong, there is a wider spread in possible patch opening angles. While the on-target binding energy ($E_{AB}$) dominates, the possible patch opening angle narrows. These behaviors not only indicate a nontrivial dependence between binding energy strength and patch locations, but they also suggest that we may not need to eliminate off-target binding energies for successful assembly.

We further examine the coupling between binding energy and patch opening angle by computing the Hessian matrix of all possible mixed second derivatives of the loss function (see Fig.~\ref{fig:energy}(b-c)) for triangle and square self-assembly optimization. We observed that the patch opening angle clearly plays a dominant role in the design optimization process, and the eigenvalues of the Hessians span several orders of magnitude. These behaviors are remarkably similar to the literature on information geometry, where the Hessian is equivalent to the Fisher information matrix in fitting a nonlinear model to data \cite{transtrum2011geometry}. The large eigenvalues correspond to the \emph{stiff} (important) parameter combinations that significantly affect the model prediction, while the small eigenvalues correspond to the \emph{sloppy} (unimportant) parameter combinations underlying emergent simplicity of models \cite{machta2013parameter}. The spread of eigenvalues across many orders of magnitude is a generic behavior of multiparameter models \cite{quinn2022information}, corresponding to a valley-like loss landscape: descent across the valley slope is very fast, but further navigation along the valley length is slow.

The conceptual goal of design optimization differs somewhat from that of statistical inference: while inference can largely disregard specific parameter values in sloppy directions, in design, the sloppy parameters are where design freedom resides. To highlight the importance of sloppy design directions, we conducted self-assembly optimizations in which only the patch opening angle, the stiffest parameter, was allowed to vary. Such optimizations have a significantly higher failure rate, taking longer to converge or converging to a region where no desired structures assemble. The sloppy directions are thus also crucial for the design process, and this work introduces Hessians as an important parameterization and visualization tool for design optimization \emph{via} gradient descent. Detailed analysis of Hessians and their spectra, global loss function landscape shapes, and the possible connection between the sloppy modes in the design process and the floppy modes in the self-assembly dynamics stand out as key questions for future work.

\section{Conclusion}
In this paper, we demonstrate how to utilize a gradient-based method as a learning tool for designing soft materials. We uncovered a nontrivial relationship between optimized design parameters and their dependency on the desired assembled structure, systematically compared inverse-design results from stabilization and self-assembly, taking into account kinetics, and provided new methods to interrogate the relationships between potential design parameters for building blocks. These insights are beneficial for any self-limiting assembly, especially in places where \emph{de novo} protein design is used for closed structure assembly, as our method can identify the precision of control needed for a given set of parameters.

\begin{acknowledgments}
The authors thank Carl Goodrich, Ella M. King, Andrei A. Klishin, and Andrea Liu for helpful discussions. This work was supported by the National Science Foundation (NSF) AI Institute in Dynamic Systems under Grant No.\ CBET-2112085, and NSF Grant DMR-2418928. The technical support and advanced computing resources from University of Hawaii Information Technology Services – Research Cyberinfrastructure, funded in part by the NSF CC* awards \#2201428 and \#2232862 are gratefully acknowledged. The computational workflow and data management for this publication was supported by the signac \cite{signac_commat} and signac-flow \cite{signac_scipy_2018} data management framework.
\end{acknowledgments}

\section*{Data Availability}
Code to reproduce the results of the manuscript has been stored at Zenodo \cite{papercode}.

\bibliography{bibliography}

\end{document}


\preprint{APS/123-QED}
\title{Patchy Particles Design: From Floppy Modes to Sloppy Dimensions - Supplementary Information}
\author{Gregory Snyder}
\author{Chrisy Xiyu Du}%
 \email{xiyudu@hawaii.edu}
\affiliation{%
 Mechanical Engineering, University of Hawai`i at Mānoa, Honolulu, HI, 96822
}%

\date{\today}

\maketitle

\section{Patch Size Determination}

\begin{figure}
  \centering
  \begin{subfigure}[t]{\linewidth}
    \captionsetup{justification=raggedright, singlelinecheck=false, position=above}
    \includegraphics[width=\linewidth]{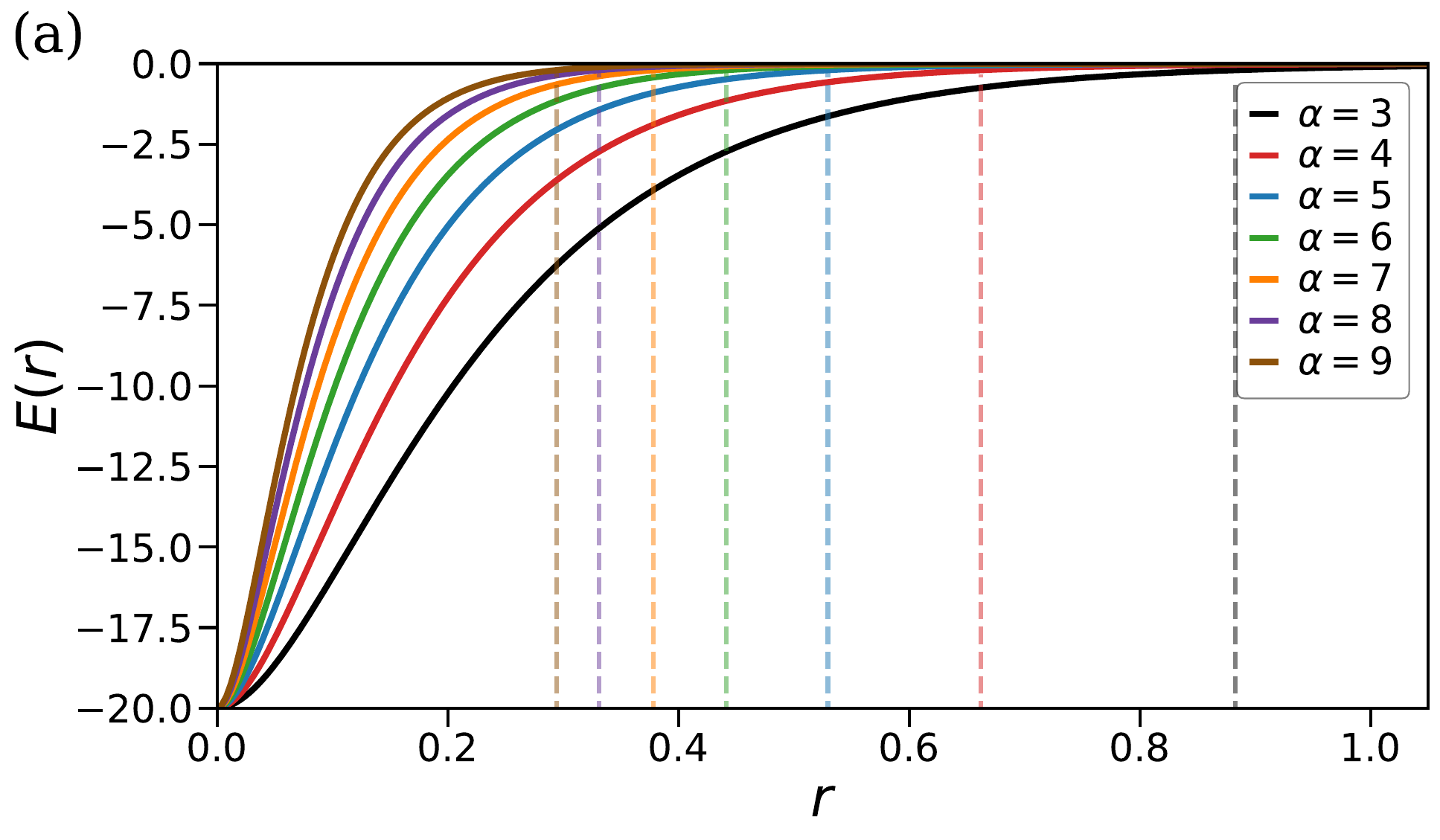}
  \end{subfigure}
  \begin{subfigure}[t]{\linewidth}
    \captionsetup{justification=raggedright, singlelinecheck=false, position=above}
    \includegraphics[width=\linewidth]{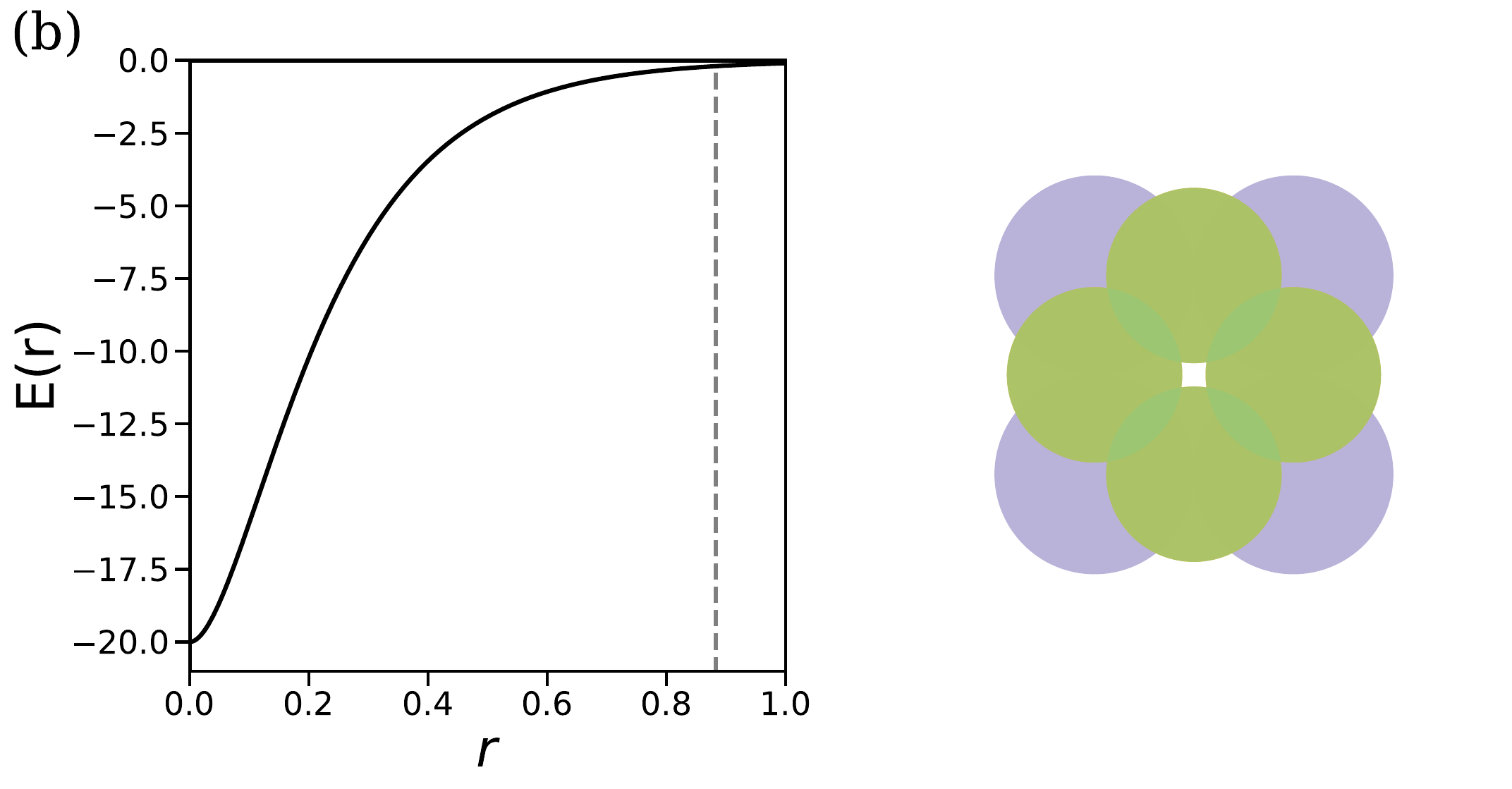}
  \end{subfigure}
  \begin{subfigure}[t]{\linewidth}
    \captionsetup{justification=raggedright, singlelinecheck=false, position=above}
    \includegraphics[width=\linewidth]{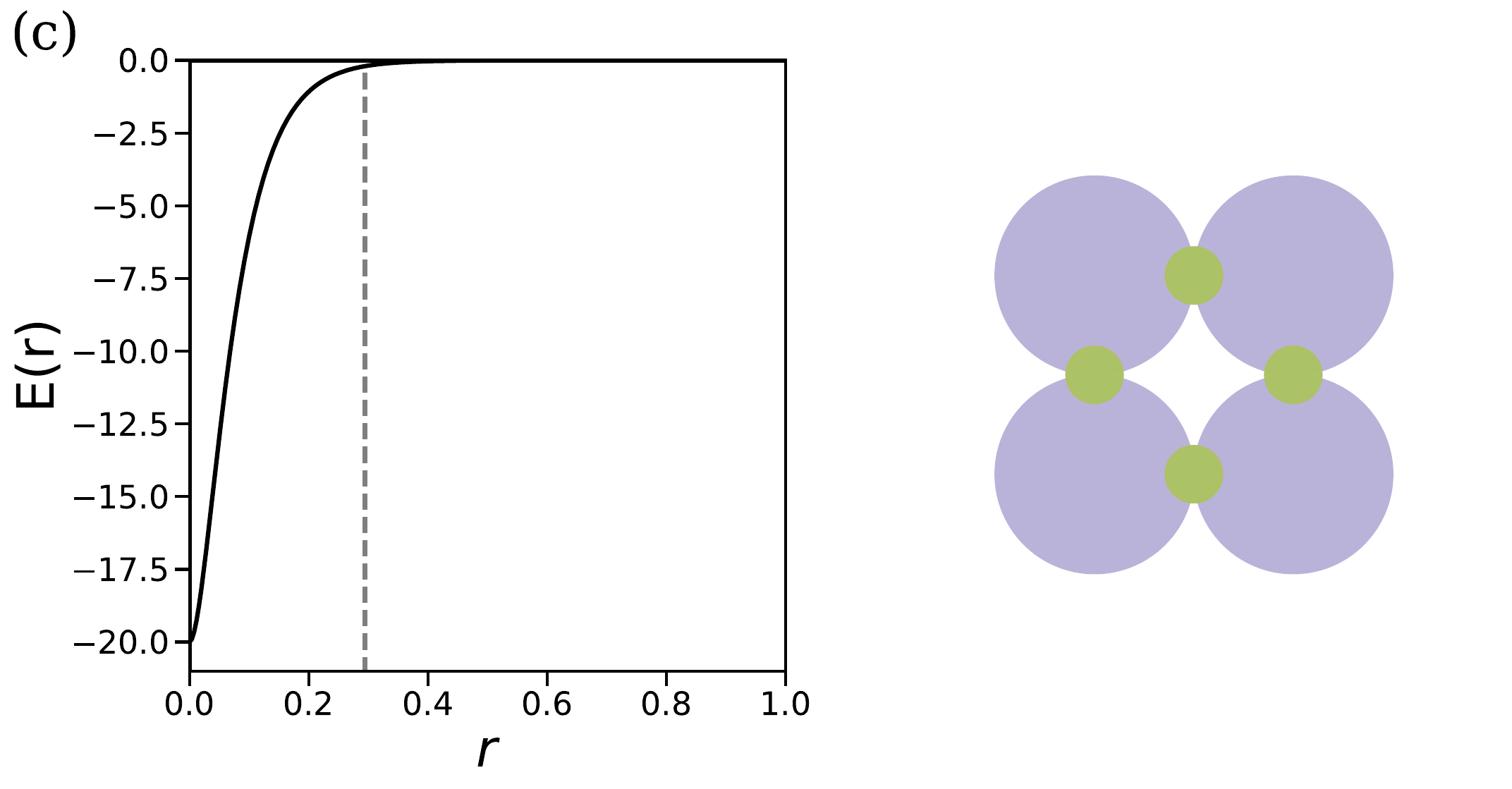}
  \end{subfigure}
  \caption{(a) Morse potential for various $\alpha$s with corresponding cut-off distances at $U_{M}(r_{\text{cut}})=0.01\epsilon$. As $\alpha$ increases, patch size decreases. (b-c) A side-by-side visualization of how changing the steepness of the Morse potential affects the patch size of the patchy particles. These figures correspond to $\alpha = [3,9]$, respectively.} 
  \label{fig:stability_snapshots}
\end{figure}

In this work, the patch size is determined by the steepness parameter of the Morse potential $\alpha$. As $\alpha$ increases, patch size decreases. In Eq. ~\ref{eqn:morse},
\begin{equation}
 U_{M}(r) = \epsilon_m(1-e^{-\alpha(r-r_0)})^2,
 \label{eqn:morse}
\end{equation}
we set $r_0=0$ so that binding is the strongest when two patches completely overlap. 
To provide an exact mapping between $\alpha$ and patch size, we set the cut-off distance $r_{\text{cut}}$ to be $U_{M}(r_{\text{cut}})=0.01\epsilon$. By treating the bond distance as the diameter of the patch, we can rearrange Eq.~\ref{eqn:morse} to solve for the patch radius below:
\begin{equation}
    a = \frac{ln(1-\sqrt{0.99})}{-2\alpha}
    \label{[eq:patch_size]}
\end{equation}

\section{Batch Size Benchmark}

\begin{figure}[H]
  \centering
  \begin{subfigure}[t]{\linewidth}
      \includegraphics[width=1.0\linewidth]{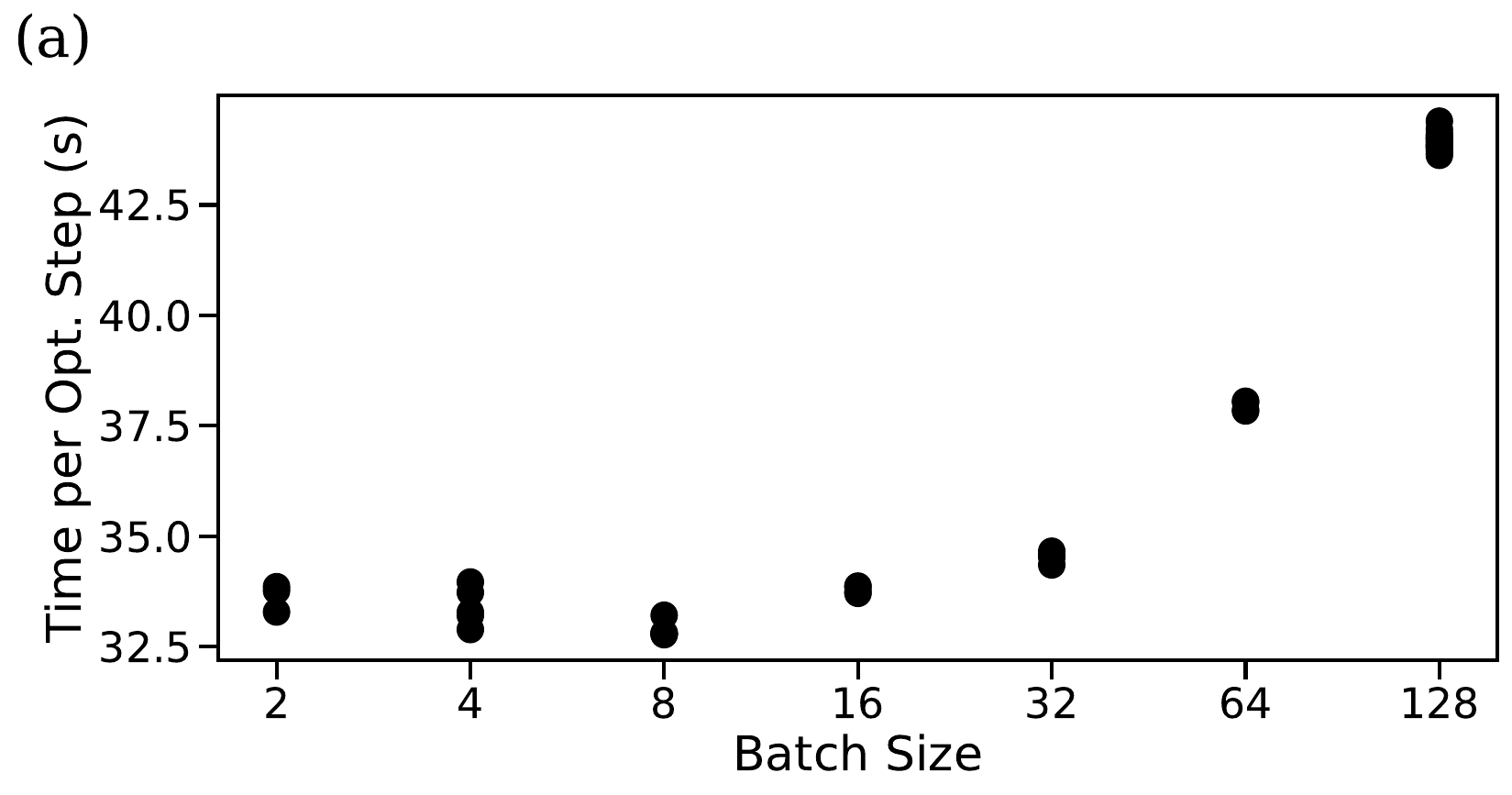}
  \end{subfigure}
  \begin{subfigure}[t]{\linewidth}
      \includegraphics[width=1.0\linewidth]{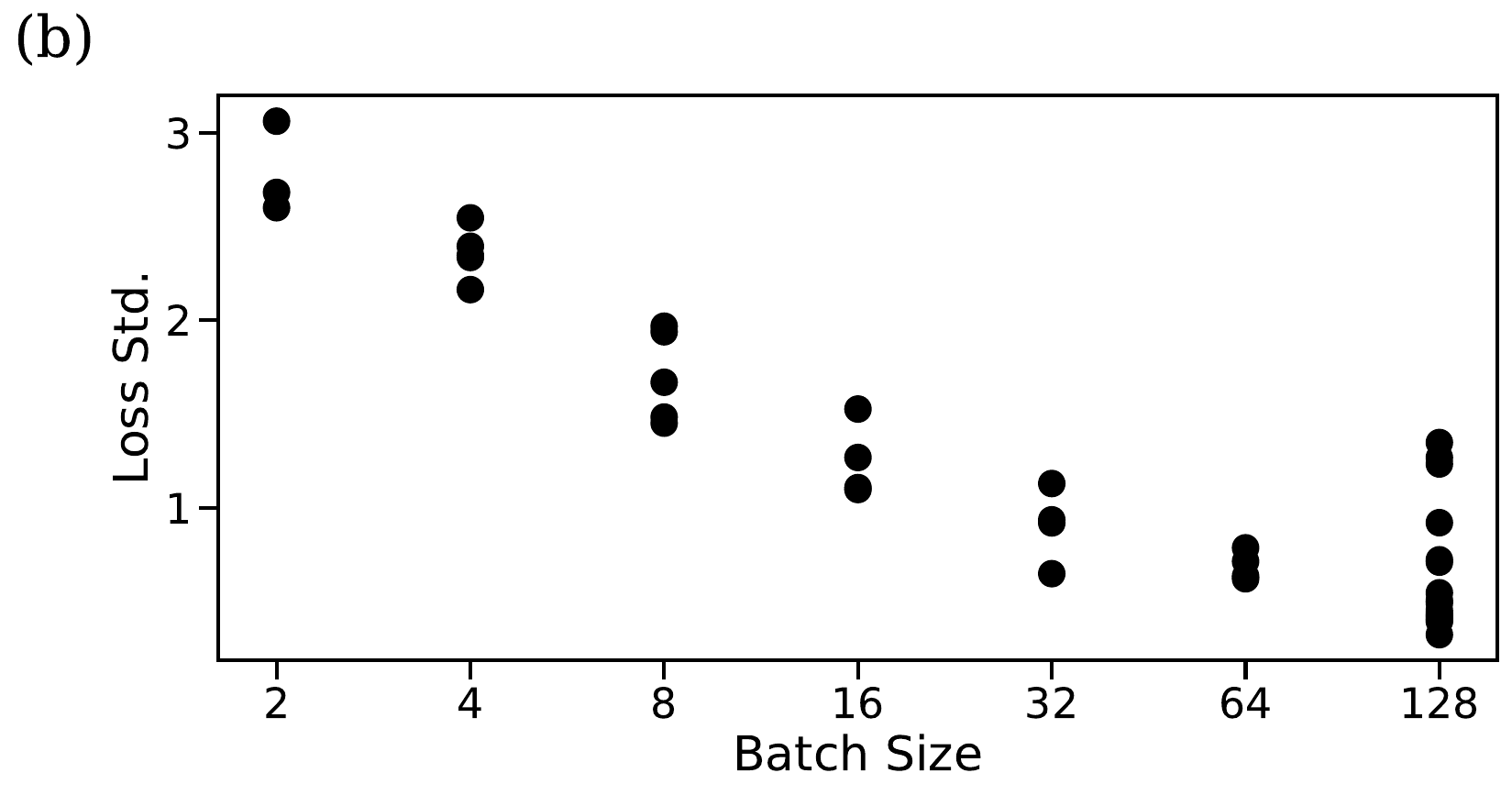}
  \end{subfigure}
  \caption{(a) Computational time (in seconds) per optimization step with respect to different batch sizes for self-assembly optimizations. (b) Standard deviation of the loss at each optimization step with respect to different batch sizes for self-assembly optimizations.}
  \label{fig:batch_study}
\end{figure}

Since our self-assembly simulation is a stochastic process, we want to average over many simulations (batch sizes) with the same parameters when taking gradients. However, as we increase batch sizes, the total optimization time will increase, and we may also encounter out-of-memory errors during back propagation. To decide the optimal batch size for our optimizations, we tested a range of batch sizes from 2 to 128. We ran five independent optimizations at batch sizes of $[2,4,8,16,32,64,128]$ on a NVIDIA H100 GPU \cite{nvidiaH100PCIe, nvidiaH100NVL}. All benchmark simulations are done on the same type of GPU for consistency, and we chose H100 for its large memory (80 Gb) as JAX-MD is very GPU-memory intensive for large batch sizes.

The benchmark runs follow the same protocols described in the method section for self-assembly optimization. We chose a patch size corresponding to $\alpha = 5$, which was arbitrarily chosen. Fig.~\ref{fig:batch_study}(a) shows the number of seconds for each optimization step at different batch sizes. We note that for batch sizes smaller than 20, there is no major time difference in an average optimization step. This results tell us that for our system, we can have a batch size of at least 16 for almost no additional time cost if the GPU memory permits. Fig.~\ref{fig:batch_study}(b) shows the standard deviation of loss as a function of batch sizes. . We note that the standard deviation plateaus after a batch size of 64, which is the size we ultimately chose for production runs.  

For stabilization optimization, we decreased the batch size to 16, as the simulations are much less noisy, allowing for quicker access to other types of GPU nodes. Since the number of H100 on the University of Hawai`i's KOA HPC \cite{uh_koa_hpc} is limited, we used NVIDIA A30\cite{nvidiaA30}, NVIDIA RTX A4000\cite{nvidiaRTXA4000}, NVIDIA Quadro RTX 5000\cite{nvidiaQuadroRTX5000}, and Tesla V100-SXM2-32GB\cite{nvidiaTeslaV100SXM2} when memory is not a huge concern.

\section{Stabilization versus Self-Assembly}

\begin{figure}
  \centering
  \begin{subfigure}[t]{\linewidth}
    \includegraphics[width=\linewidth]{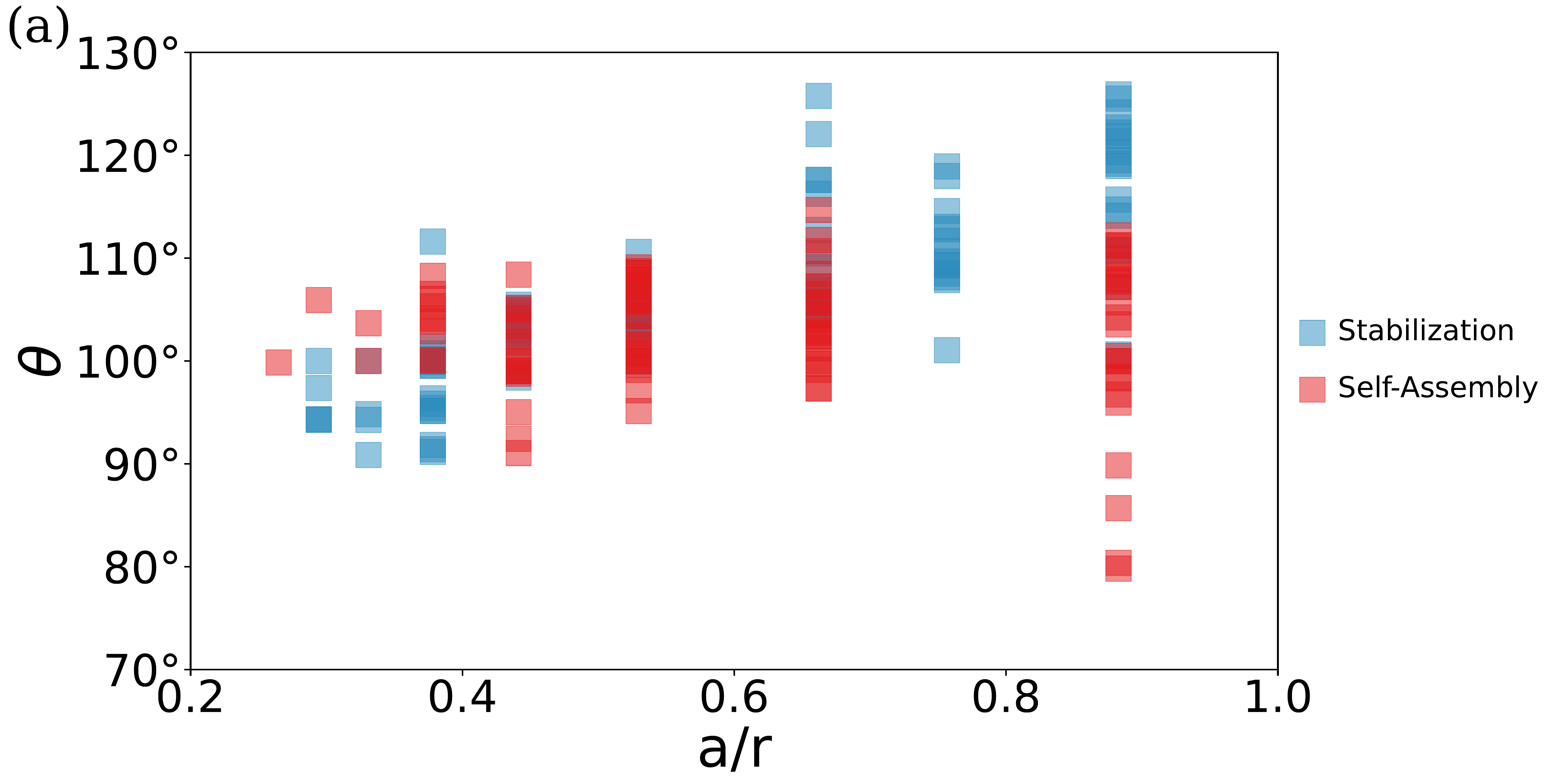}
  \end{subfigure}
  \begin{subfigure}[t]{\linewidth}
    \includegraphics[width=\linewidth]{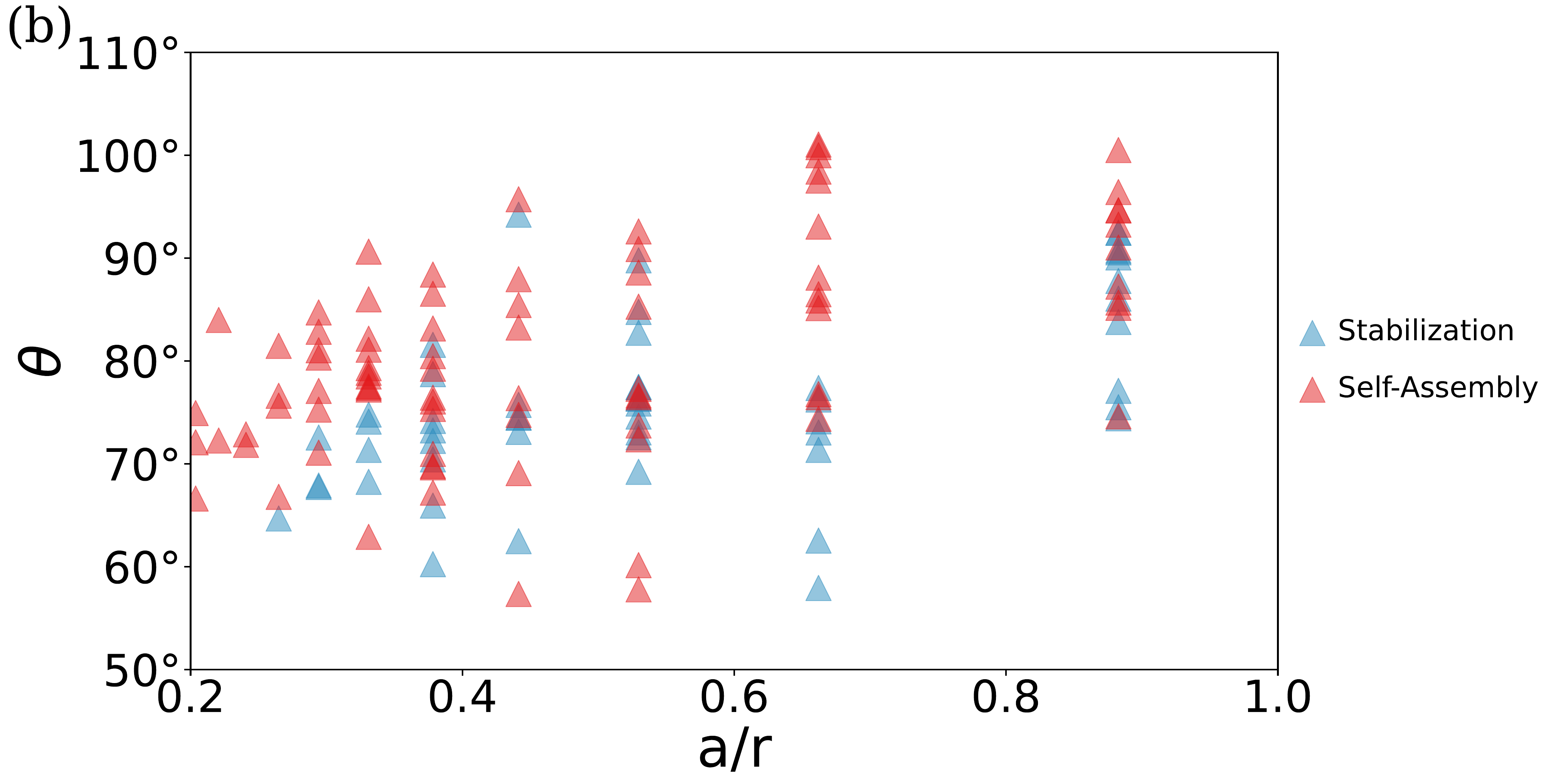}
  \end{subfigure}
  \caption{Both figures display the optimal patch opening angles collected for various patch sizes in square and triangle optimizations, respectively. The red markers represent the results from self-assembly optimizations, while the blue markers represent the results from stability optimizations. (a) An overlapping comparison of square self-assembly and square stabilization high-opening-angle morphology. (b) An overlapping comparison of the triangle self-assembly and triangle stabilization desired morphology.}
  \label{fig:optimization_compare}
\end{figure}

We provide additional comparisons between the self-assembly and stabilization optimization results for triangles and squares. For squares, we observe a notable overlap between the upper branch of the stabilization results and the self-assembly results. As the patch size decreases, the self-assembly results favor a slightly larger patch opening angle compared to stabilization; however, the number of data points is too low to be conclusive. For triangles, we plotted only the comparison between the morphology (ii) found in the stabilization optimization and the self-assembly optimization results. We note a nice overlap between the two. 

These comparisons align well with discussions in other inverse-design strategies. Often, inverse-design methods can only search for optimal parameters that stabilize the final structure, while the kinetics of self-assembly cannot be accounted for \cite{zhou2019alchemical, wei2025designing}. In our inverse-design framework, we can systematically investigate both scenarios and identify the overlapping parameters. Here, we observe that self-assembly optimizations yield a subset of stabilization optimization parameters. Since stabilization optimizations are computationally much cheaper, we could limit the scope of the optimizer when examining more complex structures before performing self-assembly optimizations in the future. Additionally, in future work, we can explore the possibility of using parameters trained on stabilization optimizations as a starting point for self-assembly optimizations. This combined approach can significantly increase the efficiency of designing building blocks using our prescribed methodologies.

\section{Binding Energy Evolution}
We provide a collection of $5$ successful, self-assembly optimization's binding energy evolution for particles with a patch ratio of $.53$. 

\begin{figure}
  \centering
  \includegraphics[width=\linewidth]{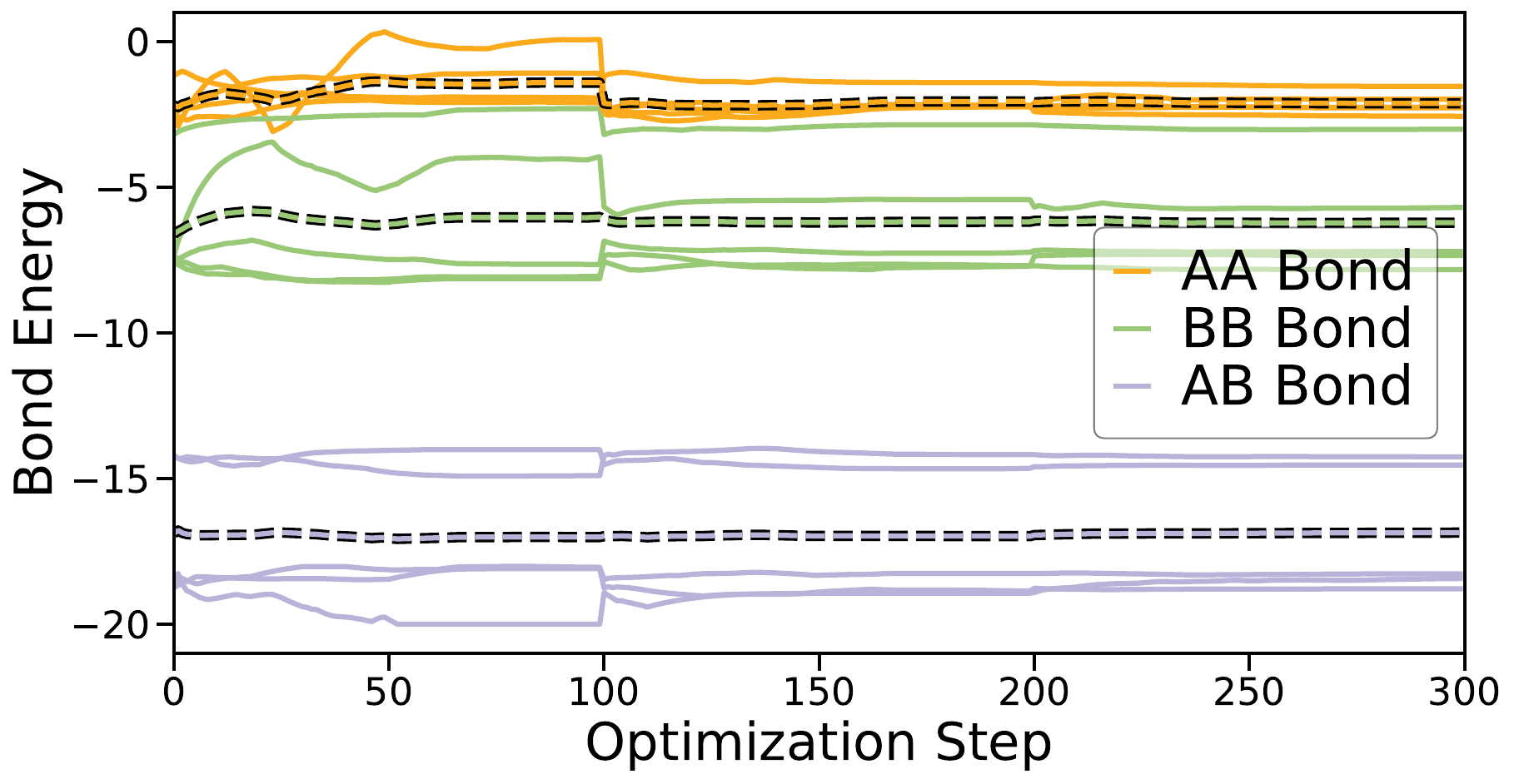}
  \caption{A collection of $5$ bond energy evolutions from successful self-assembly optimizations for particles with a patch ratio of $0.53$. The dashed lines show the average bond energies at each optimization step. }
  \label{fig:bond_evolution_sample}
\end{figure}

\section{Self-Assembly Trajectory Snapshots}
We provide a small collection of square and triangle self-assembly's final snapshots using the optimal parameters found through the gradient-based optimization for different patch sizes. These simulations provide validation for our inverse-design methods. 

\begin{figure*}
    \includegraphics[width=\linewidth]{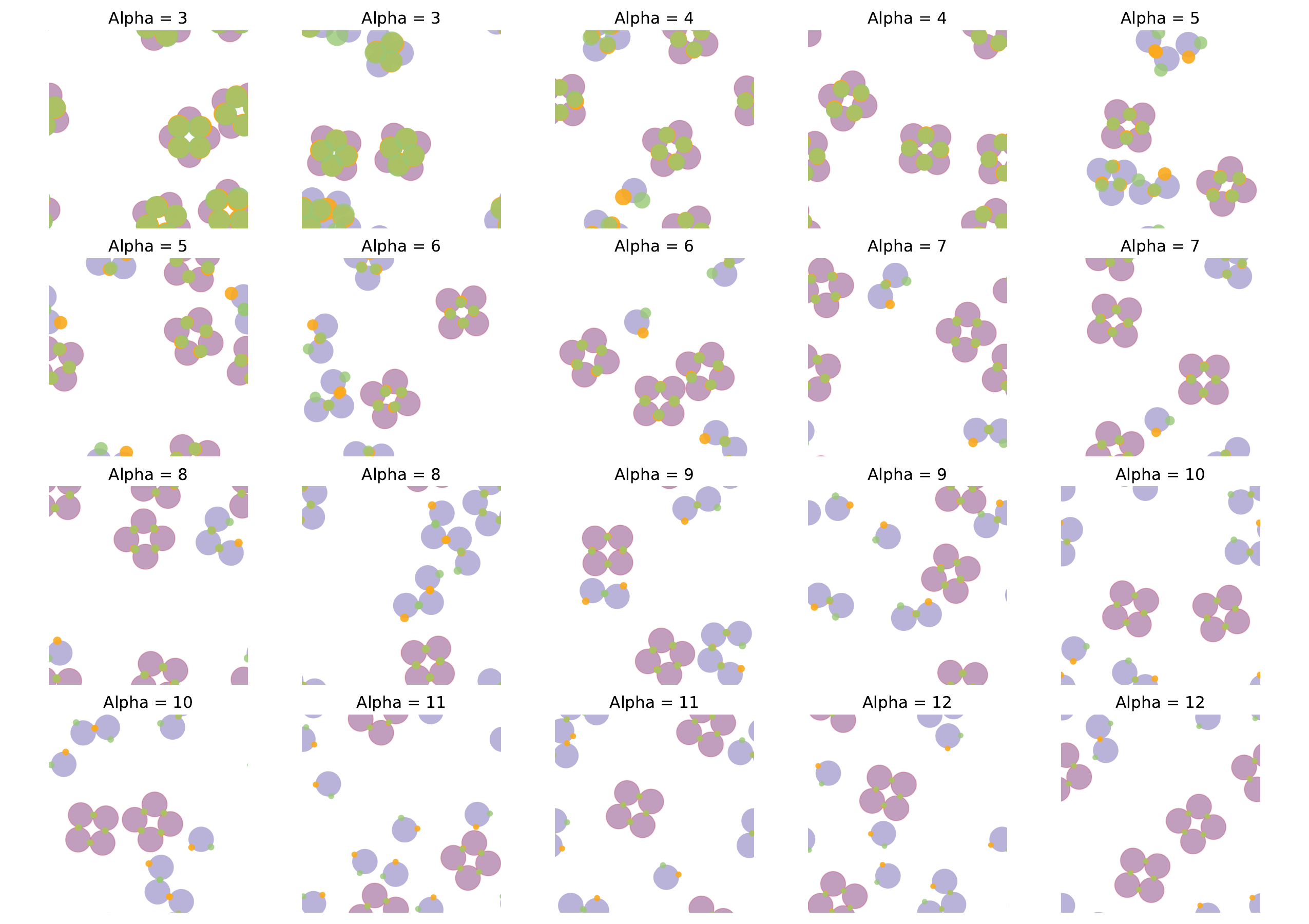}
    \caption{A collection of 20 snapshots of the final state for square self-assembly with different patch sizes. Clusters that have been identified using freud \cite{freud2020} are highlighted.}
\end{figure*}

\begin{figure*}
    \includegraphics[width=\linewidth]{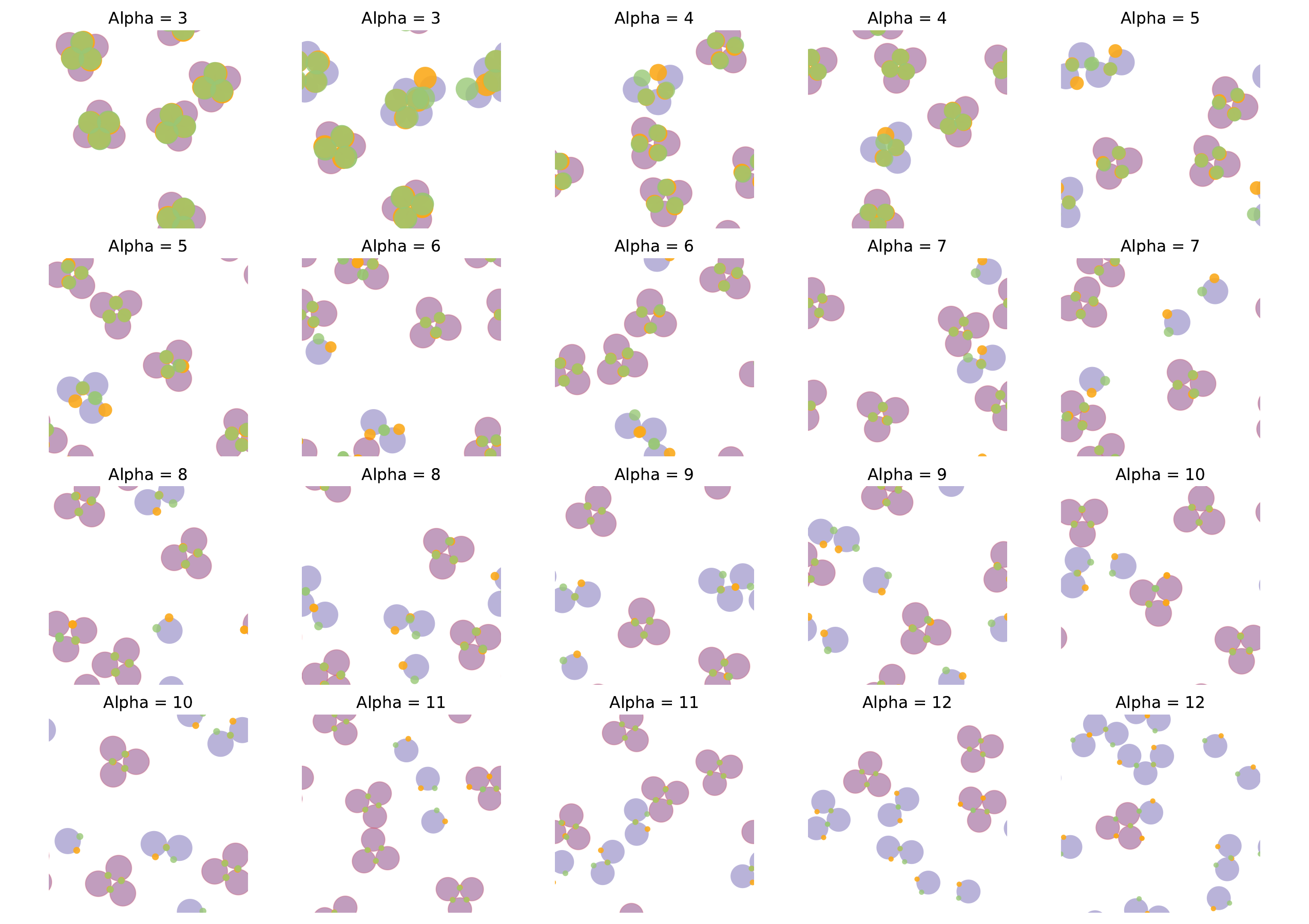}
    \caption{A collection of 20 snapshots of the final state for triangle self-assembly with different patch sizes. Clusters that have been identified using freud are highlighted.}
\end{figure*}

\section{Hessian Computation for Optimization Parameters}

We provide additional Hessian snapshots of the square and triangle self-assembly optimizations.

\begin{figure*}
  \includegraphics[width=1.0\linewidth]{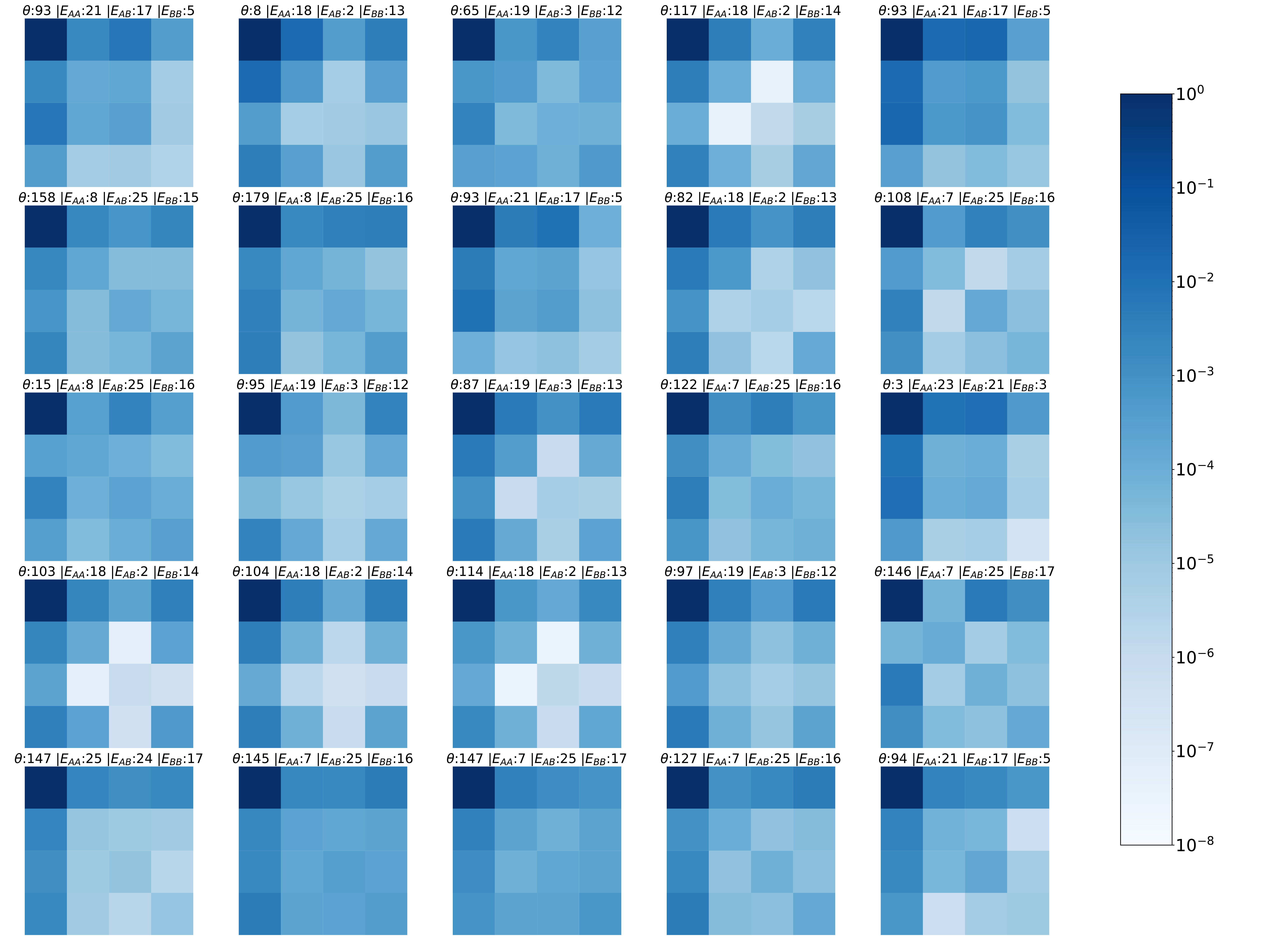}
  \caption{A collection of $25$ Hessians from snapshots of self-assembly optimizations of square structures.}
  \label{fig:additional_sq_hess}
\end{figure*}

\begin{figure*}
    \includegraphics[width=1.0\linewidth]{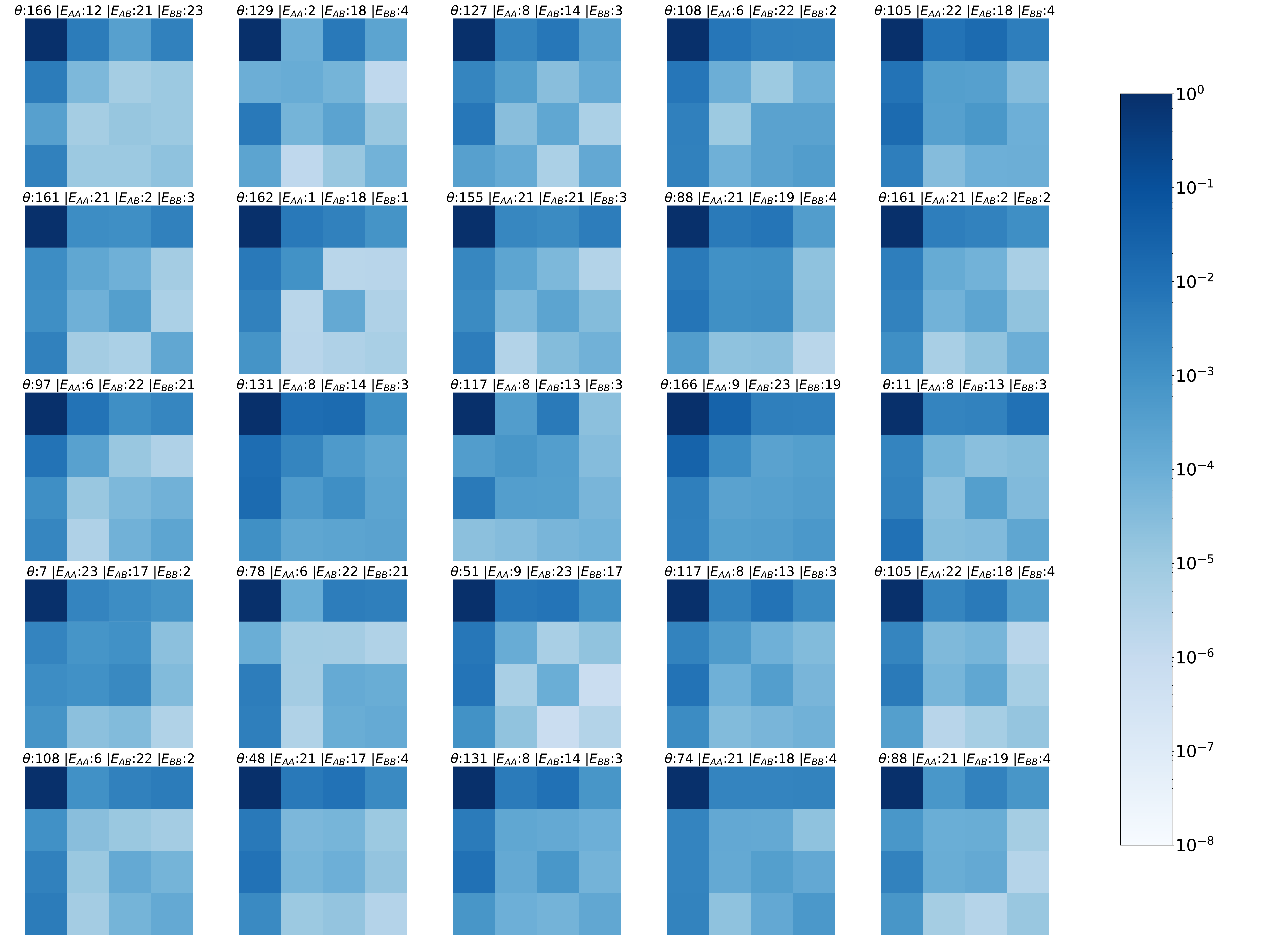}
    \caption{A collection of $25$ Hessians from snapshots of self-assembly optimizations of triangle structures.}
    \label{fig:additional_tr_hess}
\end{figure*}

\bibliography{bibliography}